\documentclass{amsart}
\usepackage{url}
\usepackage{hyperref} 
\hypersetup{backref,pdfpagemode=FullScreen,colorlinks=true}
\usepackage{color, verbatim}
\usepackage{pstricks}
\usepackage{psgo}   
 
\newtheorem{remark}[equation]{Remark}
 
\newtheorem{theorem}{Theorem} 
\newtheorem{question}{Question}

\newtheorem{lemma}[equation]{Lemma} 
 
\newtheorem{corollary}{Corollary}

\newtheorem{definition}{Definition}
\theoremstyle{definition}
\begin{document}  
\title {Simultaneous Go via quantum collapse}
\author{Yasha Savelyev} 
\begin {abstract} We construct
a symmetric, simultaneous, deterministic evolution game $SGo$, which is in a certain
mathematical sense a symmetrization of the classical board game Go. 
$SGo$ is in some ways a simpler game than Go, as Komi, Ko
and suicide rules are removed. On the other hand it has similar
dynamics and move sensitivity, enabled by certain deterministic ``quantum state'' reduction, so that state evolution is deterministic.  
Using the argument of Nash, we show that $SGo$ has
a mixed equilibrium strategy to draw on average. 
\end{abstract}
\maketitle   
\section {Introduction}    
The game Go is an ancient game of local to global geometric principles. 
Like all sequential games it suffers from
time asymmetry between the players Black and White. In Go this is addressed
by an ad hoc rule called Komi: whereby the
player White which moves second
is given a specific point advantage.  For Go, time asymmetry
also results in certain pathologies in game states, for example \emph{Ko} situation, where after a
capture of a stone the opposing player may immediately
recapture, but is forbidden by an extra rule.  The latter
sometimes leads to counterintuitive behavior, as the game
loses locality.
  
At the same time, precisely because of its local to global nature Go is ``almost time symmetric''. We make this 
precise by constructing a symmetric, simultaneous (stateful)
game $SGo$ that is in one mathematical sense
a symmetrization of Go, and on the other hand has very similar
game dynamics to $Go$. $SGo$
is a deterministic game, which is playable 
on a standard Go board, although we need a third type of
stones that we call red. It is in some ways a simpler game
than Go as it loses the Komi, Ko and suicide rules, and it has just
two rules stone placement and capture (aside from end of
game rule), but the latter rules and game states are
now a bit more complicated, as we have to deterministically
resolve simultaneous action.  

The idea for $SGo$ is loosely inspired  by quantum mechanics. In 
a sense we branch the game state whenever the moves of the two
players do not commute in classical $Go$. However, by itself
this leads to a game with uninteresting dynamics. \footnote
{It is just an example of the ``universal symmetrization'' of
a sequential game.} To get interesting dynamics,  we also have ``objective reduction of the quantum
state'', it is a kind of collapse of the wave function
arising deterministically. The term objective reduction is
coined by Penrose, see for instance
~\cite{cite_PenroseEmperor}.

Using the famous idea of Nash in
~\cite{cite_NashEquilibrium}, we show that $SGo$ has
a mixed equilibrium strategy to draw on average.  That said
finding this even approximately is likely only possible on
very small boards, due to the unreal size of
the strategy space. We can also ask if 
$SGo$ has a strategy to draw not just draw on average. 

\section{Game rules} \label{sec_Gamerules}
The game $SGo$ is initialized with an empty standard Go board,
with horizontal/vertical lines whose intersections
we just call intersections.
There are two players that we call Black
and White. As in Go, players Black and White move on a given turn by
placing a black respectively white stone on an empty intersection on the
board, but in $SGo$ they do this on the same turn without
knowledge of the other's move on this turn. This requires
modifying the placement and capture rules of Go, while
removing the Komi rule, Ko, and suicide rules. We also need to introduce
a third type of stones called red, which may be thought of as simultaneously white and black.

\subsection{Some definitions}
Given a board state $S$, that is any (at the moment no legality
restrictions) collection
of white, black, red stones on the board, let $G (S)$ denote
the graph with vertices stones, and an edge between vertices
whenever the corresponding stones are horizontally or
vertically adjacent. (Meaning they are adjacent and on the same
horizontal/vertical line.) From now on adjacency always
means this type of adjacency. In what follows the terms stones and vertices may be used synonymously. 

A \textbf{\emph{unicolor component}} $C$ is a connected
component of the full (a.k.a. induced) subgraph of $G (S)$ whose
vertices are either all the black,  or all the white stones. (Full subgraph means that all the edges of the original graph, between any
pair of vertices of the subgraph, are included.) 
In this case we call $C$ \textbf{\emph{black}},
respectively \textbf{\emph{white}}. 

As in classical Go, a connected component $C$ \textbf{\emph{has no liberties}} if it
cannot be extended (as a connected component) by placing
a stone on the board. 
We will also call $C$
a \textbf{\emph{trapped component}}.  

We define stones \textbf{\emph{adjacent}} to $C$,
as those stones whose
corresponding vertices are connected by an edge to $C$ in $G (S)$. 

Given a unicolor component $C$, 
a white/black stone on the board is called
a \emph{trapping} if it is adjacent to
a trapped unicolor component of opposite color. 

\subsection{Stone placement resolution rule of $SGo$}
If the players move to the same intersection $i$ on the board
we place a stone there whose color resolves in the following
decreasing order of priority:
\begin{enumerate}
	\item If either a white or black stone placed at $i$ is
	trapping, then a red stone is placed at $i$. We call this a \emph{trapping move}.
\item If the intersection $i$ is adjacent to
	a black stone, and neither a white stone nor red stone, a black stone is placed. If the intersection is adjacent to
	a white stone and neither a black stone nor red stone, a white stone is placed.
\item Otherwise, a red stone is placed.
\end{enumerate}
\noindent
\medskip
\begin{minipage}{0.5\textwidth} 
To the right is an example of placement resolution. On move
two both players move to E7, the resulting stone resolves to
black by our rules. Note
that the stone placed on move 4 does not resolve to white,
as it is trapping.  
\end{minipage} 
\medskip
\begin{minipage}{0.5\textwidth} 
\begin{psgoboard} [7]
\stone [1]{white} {c} {5}

\stone [1]{black} {e} {6}
\stone [2] {black} {e} {7} 
\stone [3] {red} {b} {6}
\stone [4] {red} {g} {1}
\stone {black} {e} {2}
\stone {black} {f} {2}
\stone {black} {d} {1}
\stone {white} {e} {1}
\stone {white} {f} {1}
\stone {white} {g} {2}

\end{psgoboard}
\end{minipage}


%

\subsection{Capture rule of $SGo$} \label{sec_Capture rule of SGo}
After both White and Black move on current turn numbered
$n$, using the
stone placement resolution rule above we get some board state $S$. 


\begin{definition}\label{def_}
We say that a unicolor component $C$  is
\textbf{\emph{trapped on turn $n$}} if $C$ is trapped  
and some stone adjacent to $C$, or in $C$ was placed on
turn $n$. 
A stone is called \textbf{\emph{trapped}} on turn $n$, if it
is an element of $C$ as above. We say that a red stone is
trapped by white respectively black on turn $n$, if
replacing it by a black respectively white stone it is trapped on turn $n$.
\end{definition}

A white component $C$ is resolved as
\textbf{\emph{captured}} on this turn if the following is satisfied:
\begin{enumerate}
  \item $C$ is trapped on this turn.
  \item If $C$ is adjacent to a red stone that is adjacent to another trapped component then that component is white. 
	\label{cond_unambiguous0}
	\item One of the following holds:
	\begin{enumerate}
	\item No black or red stone adjacent to $C$ is trapped by
	white on this turn. \label{cond_unambiguous}
	\item  (Suicidal capture) No stone in $C$ is created on this turn. If a stone in $C$ is adjacent to a stone trapped on
	this turn, then the latter stone is created on this
	turn and is not red. 	
	\label{cond_suicidalcapture}
\end{enumerate}
	
	\label{cond_unambiguous1}
\end{enumerate}

In the case $C$ is captured.
\begin{itemize}
	\item The stones of $C$ are removed as prisoners of Black
	as in Go. 
	\item Any red stones adjacent to $C$ resolve as black.
	(Concretely, they are replaced by black stones.)
\end{itemize}

If $C$ is a black component the process is naturally
mirrored.
The above capture rule is invoked recursively until no
captured on this turn $n$, unicolor components remain. 
\subsection{Causal reduction} \label{sec_causalreduction}
At the end of the capture procedure, we further simplify the position as
follows \footnote {It
is not logically necessary, unless we want the mathematical
relationship of $SGo$ with $Sym (Go)$ outlined in Section
\ref{sec_SGo as a symmmetrization of Go}.}. If any connected, trapped component $C$ of red stones
is adjacent only to white stones, and non of these white
stones are trapped then all stones of $C$ are replaced by
white stones. This is naturally mirrored in the case $C$ is adjacent only to black stones. The reason we call this causal reduction is outlined in Section \ref{sec_Ko situation}.
\subsection{Stage one reduction} \label{sec_Stage one reduction}
The above capture procedure followed by causal reduction is called \textbf{\emph{stage one reduction}} and the resulting board state will be denoted
by $R _{n} (S)$. 

\begin{lemma} \label{lem_} The stage one reduction is well defined,
that is independent of the order in which 
components are captured.  
\end{lemma}
\begin{proof} [Proof] 
Suppose we have two unicolor
components $C_0, C_1$ captured on the given turn. 
By assumption if a red stone is adjacent to both $C _{0}$ and $C
_{1}$ then $C _{0}$ and $C _{1}$ have the same color. 
But in that case it clearly does not matter in which order
they are captured. 

Thus suppose that no red stone is adjacent to both $C _{0}$
and $C _{1}$. If $C _{0}$ and $C _{1}$ have the same color
they cannot be adjacent, that is there is  
no pair of stones, one in	$C _{0}$ the other in $C _{1}$
that are adjacent. We now show that if they have
opposite color then they are also not adjacent. Given this, if we apply capture rule to the group $C _{0}$ and then to the
group $C _{1}$, this results in the same board state as first
applying the capture rule to the group $C _{1}$ and then the
group $C _{0}$.

We have the following possibilities. If $C _{0}, C _{1}$ both
satisfy condition \ref{cond_unambiguous}, then they are trivially not
adjacent. The same happens if $C _{0}$ satisfies the former
condition and $C _{1}$ satisfies condition \ref{cond_suicidalcapture}.
Suppose lastly that $C _{0}, C _{1}$ both satisfy
condition \ref{cond_suicidalcapture}. And suppose by contradiction that they are adjacent. Then $C _{1}$ is adjacent to a stone in $C _{0}$, which by assumptions is trapped on this turn.
Condition \ref{cond_suicidalcapture} applied to $C _{1}$ implies
that the latter stone is created on this turn. But
by assumptions no stone of $C _{0}$ is created on this turn. 
So we have a contradiction, and so $C _{0}, C _{1}$ are not
adjacent. 

To prove the general case proceed by induction. Let $S (n)$
be the sentence: for a collection $\{C _{0}, \ldots,
C _{n}\}$ of captured unicolor components, the result of stage one
reduction does not depend on the order of $C _{i}$. $S (1)$
was shown to hold. We show $S (n) \implies S (n +1)$.
Let $\{C _{0}, \ldots, C _{n+1}\}$ be a given ordered collection of
captured unicolor components, and let $\rho$ be
a permutation of the set $\{0, \ldots, n+1\}$. By the
argument above and using a sequence of adjacent
transpositions, the stage 1 reduction
applied to $\{C _{\rho (0)}, \ldots, C _{\rho (n+1)}\}$
(respecting the given order) is the stage 1 reduction
applied to  $\{C _{0}, C _{\rho' (1)}, \ldots, C _{\rho'
(n+1)} \}$ for some permutation $\rho'$ fixing 0. Using the
induction hypothesis on $\{C _{\rho' (1)}, \ldots, C _{\rho'
(n+1)}\}$ we get that the latter stage
1 reduction is the stage 1 reduction applied to the order
collection $\{C _{0}, \ldots, C _{n+1}\}$.

\end{proof}
If there are no unicolor components trapped on the turn $n-1$
or lower, the turn is resolved. Otherwise, we recursively
define higher stage reduction as follows. 

Define stage $k$ reduction
by the condition that the board state after
stage $k$ reduction denoted by $R ^{k} (S)$ satisfies:
\begin{align*}
R ^{1} (S) & = R _{n} (S). \\
R ^{k} (S) & = R _{n-k+1} (R ^{k-1} (S)).
\end{align*}
If there are no unicolor components trapped on the turn $n-k$
or lower, the capture stage is resolved after stage $k$ reduction.
\begin{remark} \label{rem_} This may appear to be fairly
complicated, but in practice the resolution of most turns is
easy to work out mentally, we will give some examples of
such ``nested entanglement states'' further ahead.  Of course
having a computer work out turn resolution automatically is
very helpful. 
\end{remark}
\subsection {End of the game}
$SGo$ game ends after both players agree to end, or
automatically after one of the players is mathematically
winning, the latter can be determined by prisoner
counts and or by geometry of the position. \footnote {In
classical Go, Chess and likewise in SGo unless one
puts certain limits, players may
conspire to extend a game indefinitely.} At this point the winner is
decided as in Go by territory and prisoner count. White's 
territory boundary consists of connected components of
white/red stones. 
Blacks territory boundary consists of connected components of
black/red stones.
Red stones are never counted as prisoners.  See Section \ref{sec_End game position} for an example.

A draw is now possible although compared to chess it must be
much less likely, (on a typical $13 \times 13$ or $19 \times
19$ go board).
\section{Examples} \label{sec_Examples}
\subsection{Capture and non-capture moves} \label{sec_Capture and non-capture moves}
Here are some examples and non-examples of capture. \\\\
\begin{minipage} {.50\textwidth}
%
On move 1 both Black and White move to C5. As this is
a capture more, C5 will resolve to white at the end of the
turn.
\end{minipage}
\begin{minipage} {.50\textwidth}
\begin{psgoboard} [7]

 \stone {black} {e} {7}
\stone [1]{red} {c} {5}

 \stone {black} {e} {6}
 \stone [\markma]{black} {c} {6}
 \stone [\markma]{black} {c} {7}

 \stone  {white} {d} {5}
\stone {white} {b} {6}
\stone {white} {b} {7}
\stone {white} {d} {6}
\stone {white} {d} {7}
\stone {black} {e} {5}
\end{psgoboard}
\end{minipage}
\medskip
\begin{minipage} {.50\textwidth}
%
Not a capture move, as the created red stone is adjacent to
an opposite color trapped component. This should make
intuitive sense since any capture would be ambiguous from
classical Go perspective.
\end{minipage}
\begin{minipage} {.50\textwidth}
\begin{psgoboard} [7]

\stone [1]{red} {c} {5}

 \stone {black} {c} {6}
 \stone {black} {c} {7}
 \stone {black} {b} {4}
\stone {black} {d} {4}
\stone {black} {c} {3}
 \stone  {white} {c} {4}

\stone {white} {b} {6}
\stone {white} {b} {7}
\stone {white} {d} {6}
\stone {white} {d} {7}
\end{psgoboard}
\end{minipage}
\medskip
\begin{minipage} {.50\textwidth}
%
White's move is a capture move, as Condition
\ref{cond_suicidalcapture} is satisfied. This is  
consistent with classical Go behavior that suicidal captures
resolve to capture. 
\end{minipage}
\begin{minipage} {.50\textwidth}
\begin{psgoboard} [7]

\stone [1]{white} {c} {5}
\stone {black} {b} {5}
\stone [1]{black} {e} {6}

\stone {black} {d} {5}
 \stone [\markma]{black} {c} {6}
 \stone [\markma]{black} {c} {7}
 \stone {black} {b} {4}
\stone {black} {d} {4}
\stone {black} {c} {3}
 \stone  {white} {c} {4}

\stone {white} {b} {6}
\stone {white} {b} {7}
\stone {white} {d} {6}
\stone {white} {d} {7}
\end{psgoboard}
\end{minipage}
\medskip
\begin{minipage} {.50\textwidth}
%
Not a capture move, now Condition
\ref{cond_suicidalcapture} is not satisfied, since the
stone placed on this turn is red.  
\end{minipage}
\begin{minipage} {.50\textwidth}
\begin{psgoboard} [7]

\stone [1]{red} {c} {5}
\stone {black} {b} {5}

\stone {black} {d} {5}
 \stone {black} {c} {6}
 \stone {black} {c} {7}
 \stone {black} {b} {4}
\stone {black} {d} {4}
\stone {black} {c} {3}
 \stone  {white} {c} {4}

\stone {white} {b} {6}
\stone {white} {b} {7}
\stone {white} {d} {6}
\stone {white} {d} {7}
\end{psgoboard}
\end{minipage}

\subsection{More complex examples} \label{sec_More complex examples}
In the first example, White 
need only play with a pure strategy (and still
be ostensibly winning). In more marginal, complex positions,
the players may need mixed strategies.  \\\\
\medskip
\begin{minipage}{.5\textwidth}
In the position on the right, on turn 1, White attempts
capture at C4, while Black attempts capture at C3. Nothing
is captured on turn 1 by our rules, and after
turn 2 the position becomes an example of a ``nested entangled state''.  
A possible continuation is indicated.   Note that F6 is not
a capture by our rules, similarly to the last example of
the previous section. But F2 is a capture move since condition
\ref{cond_unambiguous} is satisfied.
\end{minipage}
\begin{minipage}{.5\textwidth}
\begin{psgoboard}[6]
\newrgbcolor{purple}{139 0 139} 
\stone {black} {b} {3}
\stone {black} {d} {4}
\stone {white} {d} {5}
\stone {white} {d} {3}
\stone {white} {e} {4}
\stone {black} {d} {2}
\stone {white} {e} {3}
\stone {black} {f} {4}
\stone {black} {d} {6}
\stone {white} {e} {5}
\stone {black} {f} {3}
\stone {black} {f} {5}
\stone {red} {e} {2}  
\stone [3]{red} {f} {6}  
\stone [4]{red} {f} {2}  

\stone [1]{black} {c} {3}
\stone [1] {white} {c} {4}
\stone {black} {c} {5} 
\stone {black} {e} {6}
\stone  {white} {b} {5}

\stone  {black} {b} {4}
\stone  {white} {b} {5}

\stone  {black} {a} {5}

\stone  [2]{black} {b} {6}

\stone  [2]{white} {c} {6}
\end{psgoboard}
\end{minipage}
\medskip
\begin{minipage}{.5\textwidth}
At the end of turn 4, stage one resolution is depicted on the right.
\end{minipage}
\begin{minipage}{.5\textwidth}
\begin{psgoboard}[6]
\stone {red} {a} {6}  
\stone {white} {f} {2}  
\stone {black} {d} {6}

\stone {black} {e} {6}
\stone {black} {b} {3}
\stone {black} {d} {4}
\stone {white} {d} {5}
\stone {white} {d} {3}
\stone {white} {e} {4}
\stone {black} {d} {2}
\stone {white} {e} {3}

\stone {white} {e} {5}
\stone {red} {e} {2}  
\stone {white} {f} {6}  
\stone {black} {c} {3}
\stone  {white} {c} {4}
\stone {black} {c} {5} 
\stone  {white} {b} {5}

\stone  {black} {b} {4}
\stone  {white} {b} {5}

\stone  {black} {a} {5}

\stone  {black} {b} {6}

\stone  {white} {c} {6}
\end{psgoboard}
\end{minipage}
\medskip
\begin{minipage}{.5\textwidth}
Stage two resolution.
\end{minipage}
\begin{minipage}{.5\textwidth}
\begin{psgoboard}[6]
\stone {red} {a} {6}  
\stone {white} {f} {2}  
%
\stone {black} {b} {3}
\stone {black} {d} {4}
\stone {white} {d} {5}
\stone {white} {d} {3}
\stone {white} {e} {4}
\stone {black} {d} {2}
\stone {white} {e} {3}

\stone {white} {e} {5}
\stone {red} {e} {2}  
\stone {white} {f} {6}  
\stone {black} {c} {3}
\stone  {white} {c} {4}
\stone {black} {c} {5} 
\stone  {white} {b} {5}
\stone  {black} {b} {4}
\stone  {white} {b} {5}

\stone  {black} {a} {5}

\stone  {black} {b} {6}

\stone  {white} {c} {6}
\end{psgoboard}
\end{minipage}
\medskip
\begin{minipage}{.5\textwidth}
Stage three resolution.
\end{minipage}
\begin{minipage}{.5\textwidth}
\begin{psgoboard}[6]
\stone {white} {f} {2}  
%
\stone {black} {b} {3}
\stone {black} {d} {4}
\stone {white} {d} {5}
\stone {white} {d} {3}
\stone {white} {e} {4}
\stone {black} {d} {2}
\stone {white} {e} {3}

\stone {white} {e} {5}
\stone {red} {e} {2}  
\stone {white} {f} {6}  
\stone {black} {c} {3}
\stone  {white} {c} {4}
\stone  {white} {b} {5}

\stone  {black} {b} {4}
\stone  {white} {b} {5}

\stone  {black} {a} {5}

\stone  {black} {b} {6}

\stone  {white} {c} {6}
\end{psgoboard}
\end{minipage}
\medskip
\begin{minipage}{.5\textwidth}
Stage four and final resolution of turn 4.
\end{minipage}
\medskip
\begin{minipage}{.5\textwidth}
\begin{psgoboard}[6]

\stone {black} {b} {3}
\stone {white} {d} {5}
\stone {white} {d} {3}
\stone {white} {e} {4}
\stone {black} {d} {2}
\stone {white} {e} {3}
\stone {white} {e} {5}
\stone {red} {e} {2}  
\stone {white} {f} {6}  
\stone {black} {c} {3}
\stone  {white} {c} {4}

\stone  {white} {f} {2}
\stone  {white} {b} {5}

\stone  {black} {b} {4}
\stone  {white} {b} {5}

\stone  {black} {a} {5}

\stone  {black} {b} {6}

\stone  {white} {c} {6}
\end{psgoboard}
\end{minipage}
\medskip
\subsubsection{Ko situation} \label{sec_Ko situation}
Below is a Ko type situation. Recall, that we don't have a Ko rule, but it
is now unnecessary, since the board state cannot
immediately loop.  
\begin{minipage}{.5\textwidth}
White's D6 is a capture move.
\end{minipage}
\medskip
\begin{minipage} [c]{.5\textwidth}
\begin{psgoboard}[8]
 \stone {black} {c} {6}

 \stone {black} {d} {5}
 \stone {black} {e} {6}
 \stone {white} {b} {3}

 \stone {black} {f} {6}
 \stone {white} {b} {6}
 \stone {white} {c} {7}
\stone {white} {c} {8}

 \stone {white} {d} {8}
 \stone {white} {e} {7}
 \stone [\markma] {black} {d} {7}
 \stone [1]{white} {d} {6}
%
%

\stone {white} {c} {8}
 \stone [1]{black} {f} {7}
\stone  {black} {a} {7}
\end{psgoboard}
\end{minipage}
\medskip
\begin{minipage}{.5\textwidth} 
Here is a possible continuation.  
\end{minipage}
\medskip
\begin{minipage}{.5\textwidth}
\begin{psgoboard}[8]
\stone [2]{red} {d} {7}
\stone [3]{red} {c} {5}

\stone {white} {b} {3}
%
\stone [\markma]{black} {c} {6}
\stone {black} {d} {5}
\stone {black} {e} {6}
\stone {black} {a} {7}
\stone {black} {f} {6}
\stone {black} {f} {7}
\stone {white} {b} {6}
\stone {white} {d} {6}
\stone {white} {c} {7}
\stone {white} {c} {8}
\stone {white} {d} {8}
\stone {white} {e} {7}
\end{psgoboard}
\end{minipage}
\medskip
\begin{minipage}{.5\textwidth} 
At the end of turn 3 we get the following position. The
red stone at D7 is converted to white by the causal reduction rule.
To remark, this should make sense from our perspective
of $SGo$ as a ``quantum analogue'' of Go. A red stone at D7
is meant to be both black and white, but 
if it was a black stone we would have an illegal Go
position, so it must actually be white. (This is the meaning
behind the term causal reduction.)
\end{minipage}
\medskip
\begin{minipage}{.5\textwidth}
\begin{psgoboard}[8]
\stone {white} {d} {7}
\stone {white} {c} {5}

\stone {white} {b} {3}
%
\stone {black} {d} {5}
\stone {black} {e} {6}
\stone {black} {a} {7}
\stone {black} {f} {6}
\stone {black} {f} {7}
\stone {white} {b} {6}
\stone {white} {d} {6}
\stone {white} {c} {7}
\stone {white} {c} {8}
\stone {white} {d} {8}
\stone {white} {e} {7}
\end{psgoboard}
\end{minipage}

%
%
%
%

\subsection{Life and death} \label{sec:life}
Life and death is positionally very similar to Go, here is an
example. \\\\ 
\medskip
\begin{minipage} {.5\textwidth} 
The black group is alive, that is impossible to kill.
\end{minipage}
\medskip
\begin{minipage} {.5\textwidth}
 \begin{psgoboard}[7]
\stone {black} {b} {3}

\stone {black} {b} {4}
\stone {black} {b} {5}

\stone {black} {b} {7}
\stone {black} {b} {6}

\stone {white} {c} {3}

\stone {white} {c} {4}
\stone {white} {c} {5}

\stone {white} {c} {7}
\stone {white} {c} {6}
\stone {white} {a} {2}
\stone {white} {b} {2}

\stone {black} {a} {7}
\stone [1]{red} {a} {5}

\stone {red} {a} {3}
\end{psgoboard}
\end{minipage}
%
%

\subsection{End game position} \label{sec_End game position}
In the diagram below, the game has ended. Assuming there are no
prisoners, White has 3
points and Black has 19 points. 
\\\\
 \begin{psgoboard}[7] 

 \stone {black} {b} {5}
 \stone {red} {e} {6}

 \stone {red} {g} {3}
 \stone {white} {e} {1}
\stone {white} {g} {2}

 \stone {white} {e} {2}

 \stone {white} {e} {3}

 \stone {white} {e} {4}

 \stone {white} {f} {4}

 \stone {white} {f} {6}
 \stone {red} {g} {4}

 \stone {white} {f} {2}
 \stone {black} {d} {1}

 \stone {black} {d} {2}
 \stone {black} {d} {3}

 \stone {black} {d} {4}

 \stone {black} {d} {5}

 \stone {black} {d} {6}

 \stone {black} {e} {5}

 \stone {black} {f} {5}

 \stone {black} {g} {5}
 \stone {black} {b} {3}

 \stone {black} {b} {4}
 \stone {black} {b} {5}

 \stone {black} {b} {7}

 \stone {red} {c} {6}
 \stone {black} {b} {3}

 \stone {black} {a} {7}
 \stone {red} {a} {5}

 \stone {black} {a} {3}
\stone  {black} {b} {5}
\stone  {white} {c} {2}
\stone  {black} {a} {3}
\stone  {black} {a} {6}
\end{psgoboard}
\section{SGo as a symmetric simultaneous game}
We will use the formalism of automata, as it is very
flexible. The goal is to briefly introduce stateful,
symmetric, simultaneous, games.
\begin{definition} \label{def_simultaneous}
A (stateful) \textbf{\emph{simultaneous game}} $G$ with 2 players $P _{0}, P _{1}$ or Black, White, consists of: 
\begin{itemize}
\item A deterministic automaton $G$, whose set of
states is a set $S (G)$ called the set of game
states of $G$. 
\item The alphabet (that is the set of moves) is a  $\mathbb{Z} _{2}$ graded finite set  $M (G) = M _{0} (G) \sqcup M _{1} (G)  $:  interpreted as possible moves of Black and White. 
\item  Games states decompose as $$S (G) = B (G) \sqcup B (G) \times M _{0} (G) \sqcup B (G) \times M _{1} (G),$$ where $B (G)$ is some set (for example it might be the set of board states).
\item Nonempty subsets $W _{0} (G), W _{1} (G), D (G) \subset
B (G)$ of final (accepting) states with empty pairwise
intersection, which are understood as states where $P _{0}$
wins, $P _{1}$ wins or both players draw, respectively. 
\item A distinguished initial state $q _{0}  \in B  (G) $, understood as the state in which the game is initialized.
\item A partial game evolution function: 
$$E = E _{G}: S (G) \times M (G) \to  S (G),$$  which has
the properties:
\begin{enumerate}
  \item $E(s,m)$ is undefined for $s$ in one of the
	accepting states.
	\item If $s \in B (G)
	\subset S (G)$, then $E (s,m) = (s,m)$ for all $m \in
	M (G)$ s.t. $E (s,m)$ is defined. 
	\item 
	If $(s,m) \in B (G) \times M (G)$, then $E ((s,m),m') \in B (G)$ 
	if defined, and is undefined if the grading of $m$ is the
	grading of $m'$.
	\item Suppose that $m _{0} \in M _{0} (G)$, $m _{1} \in M _{1}
	(G)$, $s \in B (G) $, $E (s, m _{0})$ is defined and 
	$E (s, m _{1})$ is defined then $E (s, m _{0}, m _{1})$ is
	defined. 
	\item For $s, m _{0}, m _{1}$ as in the previous property, if $E(s, m _{0},  m _{1})$ is defined then so is $E(s, m _{1},  m _{0})$ and  
\begin{equation*}
E(s, m _{0},  m _{1}  ) = E(s, m _{1}, m _{0}).
\end{equation*} 
\end{enumerate}
\end{itemize}
\end{definition}
We recursively define:
\begin{equation*}
E (s, m _{1}, \ldots,  m _{k}   ) = E (E(s, m _{1}, \ldots,
m _{k-1}),  m _{k} ). 
\end{equation*}
We say that $G$ is \textbf{\emph{finite}} if $S (G), M (G)
$ are finite, and all game sequences have bounded length, where
a game sequence is a sequence $\{m _{i}\} _{i=1} ^{i=n}$ of moves
s.t. $$\forall 1 \leq k \leq n \;E (q _{0}, m _{1}, \ldots,
m _{k}) \text{ is defined}. $$

The above notion is more interesting when $G$ also
satisfies a certain symmetry between the two players. 
%

\begin{definition}\label{def_symmetric} We say that
a simultaneous game $G$ is \textbf{\emph{symmetric}}  if the
following conditions are satisfied. 
\begin{enumerate}
\item There is a bijection $\mathcal{S}: M (G)
\to M (G)$ taking $M _{0} (G)$ onto $M _{1} (G)$.
\item There is 
a bijection $R: B (G) \to B (G)$ s.t. for all $s \in B (G)$,
all $m _{0} \in M _{0} (G)$, $m _{1} \in M _{1} (G)$,
and for $\mathcal{S} $ as above we have:
\begin{itemize}
	\item $E (R (s), \mathcal{S} (m _{0}), \mathcal{S} (m
	_{1})) = R (E (s, m _{0}, m _{1})),$ whenever both sides are defined. Furthermore, if one side is
defined then so is the other.
	\item $R$ also induces a bijection from	$W _{0} (G)$ onto $W _{1} (G)$ and $D (G)$ onto $D (G)$.
\end{itemize}
\end{enumerate}
\end{definition}

\begin{lemma} \label{thm_simultanezation} $SGo$ is
a symmetric simultaneous game.\end{lemma}
\begin{proof} [Proof] Set $G=SGo$, with latter described in
Section \ref{sec_Gamerules}. The set $M _{0} (G)$ corresponds to
possible moves of Black and $M _{1} (G)$ the moves of White,
as in the game description.

The state space is: $$S (G) = B (G) \sqcup B (G) \times M _{0}
(G) \sqcup B (G) \times M _{1} (G),$$ where an
element of $B (G) $ is a legal board position and its
entire legal game history. By legal, we mean the history
is obeying the rules of Section \ref{sec_Gamerules}.  

The evolution map $E$ is then obviously defined. For example
if $s \in B (G) $ and $m _{0} \in M _{0} (G)$, $m _{1} \in
M _{1} (G)$ are legal moves then
$E ((s, m _{0}), m _{1}) \in B (G)  $ is 
defined to be the resolution of the
board position after applying the rules of Section
\ref{sec_Gamerules}, once Black's move $m _{0}$, and White's move $m _{1}$ are reflected on the board. 

To see that $G$ is symmetric,
let $R: B (G) \to B (G)$ be the map that interchanges
the black and white colors of the stones of the
board states, leaving red stones invariant. And let
$\mathcal{S}: M (G) \to M (G)$ interchange black and white
moves, i.e. we have a natural isomorphism $M _{0} (G) \to
M _{1} (G)$, which then induces $\mathcal{S} $. 
Then clearly the conditions of Definition
\ref{def_symmetric} will be satisfied. 
\end{proof}

\subsection{Solvability}
A theorem of Zermelo \cite{cite_Zermelo} says that for
every suitably finite sequential game  either Black or White has a winning
strategy or there is a draw strategy. Obviously $SGo$ cannot
have a winning strategy, furthermore unless we put an upper
bound on the length of a game, or on the number of position
repetitions an $SGo$ game length can be
infinite. \footnote{This is also true for Go and Chess.} So
we will assume there is such a bound, which will make $SGo$
a finite game.
\begin{question} \label{question_drawSGo}
Does $SGo$ have a strategy to draw?  
\end{question}
Of course we have simultaneous games with no strategy
to draw, the classical example is rock-paper-scissors. But
the latter has a strategy to draw on average. The latter
being a very trivial case of the existence of the Von-Neumann, Nash
equilibrium for ``matrix payoff games''. $SGo$ is not
a matrix payoff game, but Nash's proof of the equilibrium theorem readily extends to
this case. In
general:
\begin{theorem} \label{thm_} Assigning any payoff for
a win and draw, a finite simultaneous
game $G$ in the sense above has a Nash equilibrium.  
\end{theorem}
\begin{proof} [Proof] For $s \in
B (G)$ set $$M _{i} (s) = \{m \in M _{i} (G) \,|\, E (s, m)
\text{ is defined} \}.$$ 

The space of
strategies of player $P _{i}$ is then:
\begin{equation*}
\mathcal{P} _{i} = \prod _{s \in B (G)} \Delta (M _{i} (s) ), 
\end{equation*}
where $\Delta (M _{i} (s))$ is the space of
probability distributions over the finite set $M _{i} (s)$.
Since $B (G)$ is also finite, the
joint strategy space $\mathcal{P} _{0} \times \mathcal{P}
_{1}$ is a compact convex subspace of $\mathbb{R} ^{{N} } $,
for some very large $N$.

The expected payoff function on $\mathcal{P} _{0} \times
\mathcal{P} _{1}$ is easily seen to be continuous, due to
finiteness of $G$. \footnote{Possibly finiteness condition
can be relaxed.} Using this, we get that 
the multi valued function $\phi $, assigning to a joint
strategy its set of countering strategies, has a closed
graph. Furthermore, the image by $\phi $ of any point is non-empty and convex.

Then as argued by Nash ~\cite{cite_NashEquilibrium} using
Kakutani's fixed point theorem we get existence of an equilibrium point.
\end{proof}
\begin{corollary} \label{cor_} A finite symmetric simultaneous
game $G$ has a strategy to draw on average,  and in particular
$SGo$ has such a strategy.
\end{corollary}
\begin{proof} [Proof] 
Assign 1 for a win, -1 for a loss and
0 to a draw. Then it is enough to show that for any Nash
equilibrium strategy the expected payoff is 0 for both
players. 

Suppose otherwise, and suppose without loss of generality
that $$(\mathcal{D}  _{0}, \mathcal{D}  _{1}) \in \mathcal{P}
_{0} \times \mathcal{P} _{1} $$ is an
equilibrium joint strategy where Black has a positive expected
payoff. 

Let $R, \mathcal{S} $ be the symmetry
operators as in the proof of Lemma
\ref{thm_simultanezation}.
Naturally applying these operators to the joint strategy $(\mathcal{D}  _{0}, \mathcal{D}  _{1})$ we
get that there is an equilibrium joint strategy
$(\mathcal{D}  _{0}', \mathcal{D}'  _{1})$ where White has positive expected payoff. But this is absurd. 
\end{proof}

\subsection{SGo as a symmmetrization of Go} \label{sec_SGo as a symmmetrization of Go}
For any sequential game $G$ there is a canonically
associated symmetric simultaneous game $Sym (G)$, which we
may call its universal symmetrization. It is not in our
scope to discuss this in detail, and is likely well understood by experts. 
A state in this game is a set of game histories in $G$, this
should not be surprising as states of $SGo$ also can be
clearly interpreted as some sets of legal game histories in $Go$. 

From one perspective $SGo$ is $Sym (Go) $ with a 
history reduction mechanism, which is built into the rules
of $SGo$. The latter gives $SGo$ dynamic behavior similar to
$Go$. It is interesting to wonder if there is something
similar for chess.
\bibliographystyle{siam}  
\bibliography{link} 
\end{document}